\documentclass[aps,prd,reprint,nofootinbib,superscriptaddress,floatfix]{revtex4-1}
\pdfoutput=1

\usepackage[USenglish]{babel}
\usepackage[utf8]{inputenc}
\usepackage{float}
\usepackage{amssymb}
\usepackage{amsmath}
\usepackage[dvipsnames]{xcolor}
\usepackage{hyperref}
\usepackage{fullpage}
\usepackage{wasysym} 

\usepackage{bm}
\usepackage{url}

\usepackage{physics}
\usepackage{enumerate}
\usepackage{pgf}
\usepackage{graphicx}

\DeclareMathOperator{\arctanh}{arctanh}

\newcommand{\eq}[1]{Eq.~(\ref{eq:#1})}
\newcommand{\fig}[1]{Fig.~\ref{fig:#1}}
\newcommand{\sect}[1]{Sec.~\ref{sec:#1}}
\newcommand{\en}{\ensuremath{\mathcal{E}}}
\newcommand{\emin}{\ensuremath{\mathcal{E}_{\text{min}}}}
\newcommand{\lz}{\ensuremath{L_z}}
\newcommand{\lmax}{\ensuremath{L_z^{\text{max}}}}
\newcommand{\lcrit}{\ensuremath{L_z^{\text{crit}}}}
\newcommand{\carter}{\ensuremath{C}}
\newcommand{\cartercrit}{\ensuremath{C_{\text{crit}}}}
\newcommand{\cartermax}{\ensuremath{C_{\text{max}}}}
\newcommand{\lplus}{\ensuremath{L_z^+}}
\newcommand{\lcritp}{\ensuremath{L_z^{\text{crit,+}}}}
\newcommand{\lminus}{\ensuremath{L_z^-}}
\newcommand{\lcritm}{\ensuremath{L_z^{\text{crit,-}}}}

\newcommand{\fxp}{\ensuremath{f^{(4)}(x,p)}}
\newcommand{\flz}{\ensuremath{f(\en,\carter,\lz)}}
\newcommand{\fen}{\ensuremath{f(\en,\carter)}}
\newcommand{\fplane}{\ensuremath{f_{\text{eq}}}}
\newcommand{\fhern}{\ensuremath{f_{\text{H}} \pqty{\tilde{\epsilon}}}}
\newcommand{\ftildehern}{\ensuremath{\tilde{f}_{\text{H}}\pqty{\tilde{\epsilon}}}}
\newcommand{\ehern}{\ensuremath{\tilde{\epsilon}}}

\newcommand{\upot}{\ensuremath{U(\theta)}}
\newcommand{\vpot}{\ensuremath{V(r)}}
\newcommand{\vx}{\ensuremath{\tilde{V}(x)}}

\newcommand{\mbh}{\ensuremath{M_\text{BH}}}

\begin{document} 

\title{Dark matter spikes in the vicinity of Kerr black holes}
\author{Francesc Ferrer}
\author{Augusto Medeiros da Rosa}
\affiliation{Department of Physics, McDonnell Center for the Space Sciences,
Washington University, St. Louis, Missouri 63130, USA}
\author{Clifford M. Will}
\affiliation{Department of Physics, University of Florida, Gainesville,
Florida 32611, USA}
\affiliation{GReCO, Institut d'Astrophysique de Paris, CNRS, Universit\'e 
Pierre et Marie Curie, 98 bis Boulevard Arago, 75014 Paris, France}

\begin{abstract}
	The growth of a massive black hole will steepen the cold dark 
	matter density at the center of a galaxy into a dense spike, 
	enhancing the prospects for indirect detection. We study the 
	impact of black hole spin on the density profile using the
	exact Kerr geometry of the black whole in a fully relativistic
	adiabatic growth framework. We find that, despite the transfer of
	angular momentum from the hole to the halo,
	rotation increases significantly the dark matter density close to the
	black hole. The gravitational effects are still dominated by the 
	black hole within its influence radius, but the larger dark matter
	annihilation fluxes might be relevant for indirect detection estimates.
\end{abstract}
\maketitle
\section{Introduction} \label{sec:introduction}

The central regions of galaxies and clusters of galaxies are a prime target 
for searches for indirect signals of annihilations or decays of dark matter 
(DM) particles. Arguably, the brightest DM source in the sky is
the center of our Galaxy because of its close proximity and potentially large
concentration of DM. 
Although the precise DM distribution is not well constrained 
by observations, measurements of the stellar rotation curve provide robust 
evidence 
for its presence in the innermost sector of our 
Galaxy~\cite{Iocco:2015xga}.
Dissipationless DM-only simulations of halos with masses ranging from dwarf 
galaxies to rich clusters suggest that its density follows a near-universal 
cusped profile~\cite{Navarro:1995iw}, which would enhance fluxes of 
high-energy radiation originated by DM reactions. Baryonic effects could
modify the inner shape of the DM density profile, either steepening the profile
via adiabatic contraction~\cite{Blumenthal:1985qy,Gnedin:2011uj} or
softening the cusp into a core through repeated and violent oscillations in
the central potential due to energy injection from active galactic nuclei or
supernov\ae~\cite{Peirani:2006cs,Mashchenko:2006dm,Governato:2012fa,DiCintio:2013qxa}. Although neither hydrodynamic simulations~(see 
e.g.~\cite{Kuhlen:2012ft} for a review) nor dynamical 
constraints~\cite{Hooper:2016ggc,Benito:2016kyp} can determine the exact DM 
brightness of the Galactic Center, several intriguing observations of 
possible signals have fuelled a sustained interest in understanding the 
DM distribution in the central regions of the Galaxy and of large scale 
structures in general~\cite{Goodenough:2009gk,Hooper:2011ti,Fields:2014pia,Calore:2014nla,Calore:2015oya,TheFermi-LAT:2017vmf,Karwin:2016tsw}.

In addition, there is strong evidence that the Galaxy harbors a 
massive black hole ($\mbh \gtrsim 4 \times 10^6 M_\odot$)
at the center~(see e.g.~\cite{Genzel:2010zy} for a
review), which could lead to a significant increase of the DM density
in its neighborhood.    
As the black hole grows and pulls in more dark matter, the density
distribution becomes steeper. 
A Newtonian analysis with an {\em ad hoc}
treatment of particle
capture by the hole showed that a {\em spike} in the dark-matter density
is created, which causes a significant boost in the 
DM annihilation fluxes~\cite{Gondolo:1999ef}.

A fully relativistic calculation for the case of a spherical hole
was completed in~\cite{Sadeghian:2013laa} (hereafter referred to as SFW),
concluding that the Newtonian framework underestimates the dark matter
density very close to the black hole: the spike reaches significantly
higher densities, and it extends closer to the event horizon.
Using the Schwarzschild geometry, SFW found a closed form for the boundary of
the region in phase space containing bound orbits that do not cross the event
horizon, and consistently took particle capture by the black hole into
account.

On the other hand, a typical black hole is expected to be rotating rather fast. 
The spin of a supermassive hole depends on whether it gained most of its 
mass via mergers or accretion. BHs that grow mostly through disk accretion, 
adding material with constant angular momentum axis, end up spinning 
rapidly~\cite{Volonteri:2004cf}. Although the outcome of an individual 
merger event depends on the initial spin alignment, for BHs that grow through 
repeated mergers we expect a distribution of spins that peaks at
$\tilde{a} \sim 0.7$, where $a$ is the Kerr parameter related to the angular
momentum $J$ by $a\equiv J/m$~\cite{Berti:2008af,Tichy:2008du,Lousto:2009ka}
and $\tilde{a} \equiv a/G m$ is the associated dimensionless quantity (we
use units in which the speed of light $c=1$).
Millimeter VLBI observations of Sgr A${}^*$~\cite{Broderick:2008sp,Broderick:2010kx}, and the analysis of Quasi-Periodic Oscillations of hot plasma spots
in the surrounding orbiting material~\cite{Dokuchaev:2013xda} suggest a 
value of $\tilde{a} \sim 0.65$ for the spin of the supermassive black
hole in the Galactic Center, which obtains its angular momentum through 
accretion of tidally disrupted stars and gas clouds with randomly oriented
angular momenta.

As pointed out in SFW, even if the initial DM distribution is spherically
symmetric, the dragging of inertial frames induced by the rotation
of the black hole could create a flux of DM in the azimuthal direction 
proportional to the Kerr parameter $a$. Numerical investigations of dark matter
geodesics around a preexisting hole have found interesting features in the
annihilation spectrum~\cite{Schnittman:2015oma}. We here study the 
modifications in the DM spike that ensue from the rotation of the black hole by
extending the calculation in SFW to the Kerr geometry.

The main difficulty in performing this calculation for a spinning black hole 
is that, in general, no analytic expression is known for the critical Carter 
constant that separates orbits that plunge into the event horizon from those 
that do not.
We circumvent this difficulty by using a brute-force determination of 
orbit ``stability'' from its turning points. We were able to find a closed
form of the phase-space boundary for the subset of orbits that are 
contained in the equatorial plane, and we check our numerical calculations
in this particular case.

The main effect we observe is a further enhancement of the density profile in 
its innermost region, at around 5 times the gravitational radius $Gm$ of the 
black hole. Our results are obtained from an initially symmetric dark matter 
distribution and can be understood as coming from the preferential binding of 
corotating orbits to the black hole.
These particles ``feel'' a deeper potential well than the Schwarzschild case 
and are thus pulled closer to the hole, in what turns out to be the dominant 
effect over the preferential capture of counter-rotating particles by the
black hole itself.

The increase obtained is more pronounced in the equatorial plane and is more 
relevant for high black hole spins.
When $\tilde{a}=0.8$, we obtain a peak 
density that is around 70\% higher than the spherical case for DM that 
initially
follows a Hernquist profile, as shown in \fig{eqt_hern}. 
For a near-extreme black hole, the density just outside the ergosphere 
is more than an order of magnitude greater than the peak density obtained in 
the Schwarzschild case. 

We choose the Hernquist profile as a proxy for a cuspy dark matter 
distribution as suggested by dark matter only N-body simulations, which do
not include the baryonic component of the Universe. 
As suggested by hydrodynamical simulations~\cite{Maccio:2011ryn,Brook:2013lza,DiCintio:2013qxa,Schaller:2014uwa,Springel:2017tpz}, baryons could play
an important role in determining the shape of the halo, possibly leading to
the formation of a core. The results in \fig{eq_constf} 
for the constant distribution provide a good description in this case,
and show a 20\% peak density increase in the equatorial plane for 
$\tilde{a}=0.8$ compared to Schwarzschild.

Let us stress that we are assuming 
that the growth of the black hole is adiabatic, but several 
dynamical effects could affect our conclusions. For 
instance, if the black hole spirals from an initially off-center 
location~\cite{Ullio:2001fb}, if significant merger events 
occur~\cite{Merritt:2002vj}, or if gravitational
scattering off stars heats the dark 
matter~\cite{Merritt:2003qk,Gnedin:2003rj,Bertone:2005hw}, 
the density inside the spike could be considerably lower.
Recent observations for the case of Sgr A${}^*$ show 
that the density of old stars is 
flat~\cite{Buchholz:2009st,Merritt:2013}, 
or even decreasing towards the Galactic Center, which implies heating 
timescales well above $10$ Gyr. However, it has been pointed out that 
this result only holds for a small fraction of bright stars and the
evidence for the existence of a central cusp in the vicinity of the Milky
Way's central black hole is at present inconclusive (see 
e.g.~\cite{Schodel:2014wma} for a recent review).
In addition,
as discussed below, the effects of dark matter annihilations could deplete
and weaken the density profile~\cite{Vasiliev:2007vh,Shapiro:2016ypb}.
Although these effects are important, our main purpose is to understand the
general relativistic effects due to the rotation of the black hole, extending
the nonrelativistic treatment in~\cite{Gondolo:1999ef} and the relativistic
static calculation in SFW.

On the other hand, it is important to note that observables such as fluxes 
depend on integrals 
of the density profile, and the region where the enhancement occurs has a 
very small volume. Thus, 
the impact of the enhancement on integrated effects will be small,
but should still 
be taken into account in model building~\cite{Fields:2014pia,Arina:2014fna,Gammaldi:2016uhg}. Moreover, a significant number of
supermassive black holes in Active Galactic Nuclei are known to be rapidly
spinning~\cite{Reynolds:2013qqa}, which could potentially enhance their
contribution to the isotropic gamma-ray background~\cite{Belikov:2013nca}.

The rest of this paper provides the details supporting these conclusions. In
\sect{methods} we describe the phase space available to bound 
orbits around a Kerr black hole. In \sect{constf} we obtain the DM 
density profile 
resulting from an initially constant phase-space distribution, and in
\sect{hernquist} we obtain the profile from an initally cuspy Hernquist 
distribution.
\sect{discussion} discusses the implications of these DM distributions
for the gravitational environment around the black hole and for the fluxes
of radiation from the DM spike.
Concluding remarks are presented
in \sect{conclusions}.

\section{Black hole growth in a dark matter halo} \label{sec:methods}
We begin by reviewing the analysis of how the adiabatic 
growth of a black hole modifies the density and velocity dispersions
within a preexisting dark matter halo. We follow the general relativistic
approach in SFW, which extended the Newtonian treatment 
in~\cite{Young:1980apj,Quinlan:1994ed,vanderMarel:1998tm} (see 
also~\cite{binneytremaine2008}).

Our starting point is the relativistic phase space distribution \fxp\ 
describing a system of dark matter particles of rest mass 
$\mu$~\cite{Walker:1936,Bernstein:1988bw,1968ApJ...153..643F}, 
normalized so that when integrated over phase space it gives the
total mass of the halo. 
The mass current four-vector can be written in terms of the distribution
function as:
\begin{equation}
  \label{eq:define_J}
	J^{\mu}(x)=\int{ \fxp u^{\mu}\sqrt{-g} \dd[4]{p}},
\end{equation}
where $u^{\mu} = p^{\mu}/\mu$ is the four-velocity, and $g \equiv 
\det(g_{\mu \nu}(x))$ is the determinant
of the metric\footnote{Note that we use contravariant components 
$\dd{p^\mu}$ to define the volume element in momentum space, 
but the same results can be obtained by taking covariant components
$p_\mu$ as the argument of the distribution function.}.
Knowledge of the mass current four-vector will allow us to find the 
dark matter density:
from the definition $J^{\mu}=\rho u^{\mu}$ and the fact that 
$u^{\mu} u_{\mu}=-1$, we obtain the mass density as
measured in a local freely falling frame as
$\rho=\sqrt{-J_\mu J^\mu}$.

To calculate $J^\mu$ we need to know the distribution function $\fxp$ and the 
boundary of integration over momentum space. Both tasks are greatly simplified
if instead of $p^\mu$ we use invariant constants of the motion to write
down the distribution function and the volume element.
We review below this change of variables for the case of a Kerr black hole 
background following the discussion in SFW.

We use Boyer-Lindquist coordinates to express the Kerr line element 
for a hole of mass $m$ (with $c=1$):
\begin{align}
  \label{eq:line_element}
	\dd{s^2} = &-\left(1-\frac{2Gmr}{\Sigma^2}\right)\dd{t^2}
	+\frac{\Sigma^2}{\Delta}\dd{r^2} \nonumber \\
	   &+\Sigma^2 \dd{\theta^2} 
	   -\frac{4Gmar}{\Sigma^2}\sin^2\theta \dd{\phi} \dd{t} \nonumber \\
  &+\left(r^2+a^2+\frac{2Gmra^2\sin^2\theta}{\Sigma^2}\right)\sin^2\theta 
	\dd{\phi^2},
\end{align}
where $a$ is the Kerr parameter defined above, and 
we have introduced the functions $\Delta=r^2+a^2-2Gmr$ and 
$\Sigma^2=r^2+a^2\cos^2\theta$. 

The orbits of dark matter particles of rest mass $\mu$ in this geometry admit 
four constants of the motion: the energy per unit mass, \en, and angular 
momentum per unit mass, \lz, that are conserved 
because the metric is
stationary and axisymmetric; 
the mass-shell condition; and the so-called Carter 
constant per unit (mass)${}^2$, \carter~\cite{Carter:1968rr}:
\begin{align}
  \label{eq:change_var}
  	\en&\equiv-u_t=-g_{tt}u^t-g_{t\phi}u^{\phi},  \nonumber \\
   \lz&\equiv u_{\phi}=g_{\phi\phi}u^{\phi}+g_{t\phi}u^t, \nonumber \\
	\mu&=\sqrt{-p_{\mu}p^{\mu}}, \nonumber  \\
	\carter&\equiv \Sigma^4(u^{\theta})^2+\frac{\lz^2}{\sin^2\theta}
	+a^2\cos^2\theta(1-\en^2). 
\end{align}
We use a definition of the Carter constant that, in the spherically symmetric
Schwarzschild limit ($a\rightarrow 0$), reduces to the square of the 
conserved total angular momentum per unit mass, $L$: $\carter \rightarrow
L^2 = u_\theta^2 + u_\phi^2/\sin^2 \theta$.

Given these definitions and using $p^{\mu}= \mu u^{\mu}$, we calculate the 
(inverse) Jacobian
\begin{equation}
  \label{eq:jacobian}
  \mathcal{J}=\left\rvert \frac{\partial(\en,\carter,\lz,\mu)}{\partial(p^t,
	p^r,p^{\theta},p^{\phi})}\right\rvert
	=\frac{2\Sigma^4 \Delta |u_r| |u^{\theta}|\sin^2\theta}{\mu^3}.
\end{equation}

The last step is to write the necessary four-velocity components appearing in 
the Jacobian in terms of the constants of the motion. These are
\begin{align}
  \label{eq:4velocity}
  u^{\theta}&= \pm\frac{1}{\Sigma^2} \sqrt{\upot} \nonumber \\
  u_r&=\pm \frac{r^2}{\Delta}\sqrt{\vpot},  
\end{align}
where
\begin{align}
	\label{eq:upotential}
	\upot \equiv \carter-\frac{\lz^2}{\sin^2\theta}
	-a^2(1-\en^2)\cos^2\theta,
\end{align}
and
\begin{align}
  \label{eq:potential}
  \vpot\equiv&\left(1+\frac{a^2}{r^2}+\frac{2Gma^2}{r^3}\right)\en^2 
	-\frac{\Delta}{r^2}\left(1+\frac{\carter}{r^2}\right) \notag \\
	&+\frac{a^2 \lz^2}{r^4} 
	-\frac{4Gma\en \lz}{r^3}.
\end{align}
The presence of the $\pm$ signs in \eq{4velocity} implies that 
the $r$ and $\theta$ components of the current $J_{\mu}$ vanish, 
as they come with an absolute value in the Jacobian and we must integrate 
over both positive and negative values of $u_r$ and $u_{\theta}$ in 
\eq{define_J}, since they are equally likely to be positive or negative for a
given set of values for \en, \carter\ and \lz. For the other two components, 
we must put in an extra factor of 4 to allow for the integration over the 
positive 
and negative values of both four-velocity components. Note that, as pointed
out in SFW, $J^\phi$ will not vanish even for spherically symmetric dark 
matter distributions, \fen, since the last term in \eq{potential} is linear
in \lz.  
This effect, related to the dragging of inertial
frames caused by the rotation of the black hole, will influence
the dark matter spike around the black hole as we further elaborate below.

Concerning the distribution function, 
we are assuming that all of the dark matter particles have the same mass 
$\mu$. 
Then we are allowed to write 
$\fxp=\mu'{}^{-3} \flz \delta(\mu'-\mu)$, which
will allow us to perform the integration over $\dd{\mu'}$.

Putting everything together, we can write the nonzero components of the 
mass current four-vector as
\begin{subequations}
  \label{eq:final_current}
\begin{align}
	J_{t}(r,\theta)&=\frac{-2}{r^2\sin\theta}
	\int  \dd{\en}\dd{\carter} \dd{\lz}
	\frac{\en \flz}{\sqrt{\vpot}\sqrt{\upot}}, \\
	J_{\phi}(r,\theta)&=\frac{2}{r^2\sin\theta}
	\int  \dd{\en}\dd{\carter}\dd{\lz}
	\frac{ \lz \flz}{\sqrt{\vpot}\sqrt{\upot}}, 
\end{align}
\end{subequations}
where we used $u_t=-\en$ and $u_\phi =\lz$ to show the covariant components.

Introducing $\Omega \equiv J_{\phi}/J_t$, the density in the rest frame of
the distribution, $\rho=\sqrt{-g^{\mu\nu} J_{\mu}J_{\nu}}$, 
reads\footnote{We correct a typo in Eq.~(3.11) in SFW.}:
\begin{equation}
  \label{eq:density}
  \rho= \left|J_t\right| \sqrt{\frac{g_{\phi\phi}-2g_{t\phi}\Omega 
	+g_{tt}\Omega^2}{\Delta\sin^2\theta}}.
\end{equation}

To actually evaluate \eq{final_current}, we still need to specify the shape of 
the integration region over \en, \carter\ and \lz, as well as the distribution 
of the trajectories of dark matter particles \flz. The first task is quite
involved and is one of the main results of this work. We give several general
considerations here that are further developed in the following sections.

\subsection{Region of integration in \en-\carter-\lz space} 
\label{sec:phase_space}

If we consider only the contribution of bound particles to the density, there 
are three constraints we must apply to our phase space: the energy $\en$ is 
bounded above by 1, and we must also have $\vpot \geq 0$ and 
$\upot \geq 0$. Also, as 
pointed out in previous work \cite{Gondolo:1999ef,Sadeghian:2013laa}, we must 
remove orbits that plunge into the black hole. For the case of a Schwarzschild
black hole, this capture condition can be worked out analytically. The 
criterion obtained in SFW is that, given an energy, there is a critical value 
of the angular momentum $L^{\text{crit}}$ 
below which all orbits are captured.

\begin{figure}[h]	
	\begin{center}
		\input{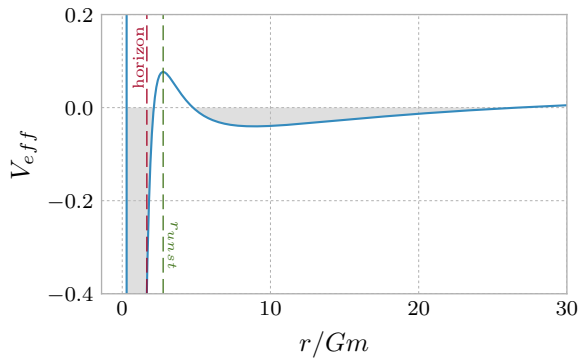}
	\end{center}
	\caption{A plot of $V_{\text{eff}} \equiv -V(r)$ vs $r/Gm$ for
	a Kerr black hole with $\tilde{a}=0.75$, 
	$\en=0.97$, $\carter=12 \pqty{Gm}^2$, and $\lz=2 Gm$ . This is 
	equivalent to the usual 
	effective potential of classical mechanics. 
	The two ``bound orbits'' are clearly seen ranging along the shaded 
	regions. The bottom of the leftmost peak 
	is not shown, as it is much lower than the normal bound orbit.}
  \label{fig:sample_potential}
\end{figure}

For the Kerr metric, no such constraint has been derived. For given values of 
the conserved quantities (\en, \carter, \lz), we now have in general two sets 
of ``bound orbits'': one orbit with two 
turning points beyond the horizon, and one with a subhorizon turning point \cite{o2014geometry}. These can be seen extending through the shaded regions
in \fig{sample_potential}, which shows
the effective potential $V_{\text{eff}} \equiv -\vpot$. In between the
two shaded regions corresponding to the two orbits, the effective potential
has a maximum at $r_{\text{unst}}$.

\begin{figure}[h]
 	\begin{center}
		\input{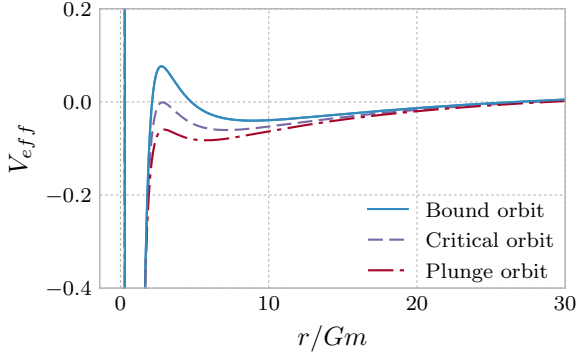}
	\end{center}
	\caption{A plot of $V_{\text{eff}}(r)$ vs $r/Gm$ with 
	fixed \en, \lz.   
	Varying the Carter constant separates bound and plunge orbits.
	The solid blue line depicts the same orbit shown in 
	\fig{sample_potential}. Decreasing \carter\ we reach the dashed 
	magenta line showing the critical orbit with \carter=\cartercrit,
	which is equal to $\cartercrit=10.3 \pqty{Gm}^2$ in this case.
	For smaller \carter\ the orbits plunge into the hole, as depicted
	by the dot-dashed red line with $\carter=9 \pqty{Gm}^2$.}
  \label{fig:crit_orbit}
\end{figure}

We can thus exclude the plunge orbits on a case by case basis using the 
following criteria: 
\begin{enumerate}
	\item The value of $r$ at which we are evaluating the current must be 
		to the right of the unstable orbit $r_{\text{unst}}$. 
	\item The point $r_{\text{unst}}$ must also be in the forbidden region,
		$V_{\text{eff}} \pqty{r_{\text{unst}}} = -
		V\pqty{r_{\text{unst}}}>0$, so that there is a potential
		barrier between the orbit of interest and the horizon. 
		This will ensure that the orbit does not cross the 
		horizon and become trapped.
\end{enumerate}
In this way we exclude all plunge orbits and all unphysical orbits with a
subhorizon turning point, as illustrated in \fig{crit_orbit}. 
Unfortunately, it is not possible in general to find analytic expressions of
the region in $(\en, \carter, \lz)$--space where these two criteria are met,
so our code implements the capture condition numerically.
Nevertheless, useful insight can be gained by focusing on a subset of the 
orbits. The case of nonrelativistic particles with $\en \approx 1$ was
studied by Will~\cite{Will:2012kq}, who found an approximate analytic 
expression for the critical value of \carter. 
We consider in \sect{equator} all bound orbits that are contained in the 
equatorial plane. In this case
$\lz = \pm \sqrt{\carter}$, which makes the problem tractable and also 
provides a useful check of the full numerical calculation.

\subsection{Adiabatic approximation} \label{sec:adb}

Now that the available phase space has been determined, at least implicitly,
we turn to \flz. 
Our starting point is a known initial
nonrelativistic dark matter distribution without a central black hole,
$f'(E,L^2,L_z)$. For simplicity, we will assume our initial distribution to 
be spherically symmetric, generating a potential $\Phi (r)$. 

As reviewed in 
appendix~\ref{ap:adiabatic_f}, the adiabatic growth of the black hole 
preserves the form of the distribution function, $\flz = f'(E,L^2,L_z')$.
Here, $E$, $L^2$ and $L_z'$ are 
obtained from the Kerr constants of the motion by noting that each particle
responds to the slow change in the gravitational potential 
by altering its energy $E$ and angular momentum 
$L$ and \lz, in such a way that the action variables $I_r$, $I_\theta$ 
and $I_\phi$ are kept fixed~\cite{Sadeghian:2013bga,Young:1980apj}.

For an orbit in the initial nonrelativistic dark matter distribution, these 
are
\begin{align}
  \label{eq:adb_invars_nr}
	I_r'(E,L) &\equiv \oint \dd{r} \sqrt{2E-2\Phi-L^2/r^2}, \nonumber \\
	I_{\theta}'(L,L_z')&\equiv \oint \dd{\theta} \sqrt{L^2-\frac{L_z'^2}{\sin^2\theta}}=2\pi (L-|L_z'|), \nonumber \\
	I_{\phi}'(L_z')&\equiv \oint \dd{\phi} L_z'=2\pi L_z'.
\end{align}
For a bound orbit in the Kerr geometry,
\begin{align}
  \label{eq:adb_invars}
	I_r(\en,\carter,\lz)& \equiv \oint u_{r} \dd{r} = \oint \dd{r} 
	\frac{\sqrt{\vpot}}{1+\frac{a^2}{r^2}-\frac{2Gm}{r}}, \nonumber \\
	I_{\theta}(\en,\carter,\lz)& \equiv \oint u_{\theta}\dd{\theta} 
	=\oint \dd{\theta} \sqrt{\upot},  \nonumber\\
	I_{\phi}(\lz)&\equiv \oint u_{\phi}\dd{\phi}=2\pi L_z.
\end{align}

The equality $I_\phi = I_\phi'$ implies $L_z'=L_z$. The conservation of this 
component of the angular momentum should be expected as both the initial and 
the final states have axial symmetry.

Given a set (\en, \carter, \lz) for a Kerr orbit, 
together with $\lz=\lz'$, we can determine $L$ via 
\begin{displaymath}
	L = \vqty{\lz'} + \frac{I_{\theta}(\en,\carter,\lz)}{2\pi}.
\end{displaymath}
Given this value of $L$, we can obtain $E$ by equating the radial actions.
Since the 
radial integral in (\ref{eq:adb_invars_nr}) cannot in general be solved 
analytically, we find the energy $E$ using the bisection method.

\section{Density profile for a constant distribution function} 
\label{sec:constf}

The simplest possible example that can be considered is that of a constant 
distribution function. Although somewhat unrealistic, it can be seen as a toy
model for describing the stars close to a black hole forming in the core of an approximately isothermal system. 
This case does not require any adiabatic matching, but is 
still useful for building intuition, as the current-density will be directly 
related to the total phase space volume. 

\subsection{Restriction to equatorial orbits}
\label{sec:equator}

We first consider the subset of particles following planar orbits,
which are only possible on the equatorial plane.
In this case, the calculations can be 
carried out analytically to a large extent. 
Moreover, the result will be a lower 
bound on the final dark matter density on the plane, because 
the calculation omits 
nonplanar orbits that cross $\theta=\pi/2$. 

We will begin by 
setting $\flz=\fplane \delta \pqty{u_{\theta}}$ and focusing on
$\theta = \pi/2$. 
This allows us to perform the integrals in \eq{final_current}
over \carter\ and \lz, obtaining
\begin{align}
  \label{eq:current_equator}
	J_t&=\frac{- 8 \fplane}{\sqrt{Gm r}} 
	\int\limits_{\emin (r)}^1 \dd{\en} \en \mathfrak{I}_t (\en), \nonumber \\		J_\phi&=\frac{ 8 \fplane \sqrt{G m}}{\sqrt{r}} 
	\int\limits_{\emin (r)}^1 \dd{\en} \mathfrak{I}_\phi (\en). 
\end{align}
The explicit forms of the functions $\mathfrak{I}_{t,\phi}$ representing
the \carter-\lz\ integration are given in \eq{ifunctions}. The \en\ dependence
in these functions is implicit in the quantities $\lz^\pm$, which are 
the two roots of the equation $\vpot=0$ given in \eq{lmax}, 
and \lcrit, the critical angular 
momentum for capture by the black hole. 
For a corotating planar orbit, $\lplus \geq \lcritp >0$, while for a
counter-rotating orbit $\lminus \leq \lcritm < 0$.
The currents
are thus naturally separated in a counter-rotating and a corotating part
corresponding to the integration domain $\Delta \lz$ in \lz:
\begin{displaymath}
	\Delta \lz = \pqty{\lminus,\lcritm} \cup \pqty{\lcritp,\lplus}.
\end{displaymath}
The minimum energy $\emin(r)$ is found by setting 
$\lcrit{}^\pm(\en)=L_z^{\pm}(\en)$. 

\begin{figure}[h]
 	\begin{center}
		\input{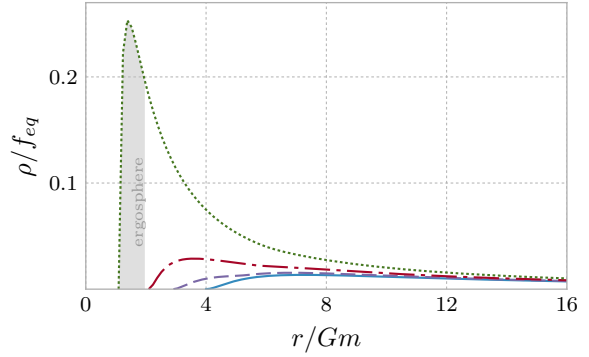}
	\end{center}
  \caption{Density profiles  
	for a distribution of equatorial orbits. The 
	density increases as we vary the Kerr parameter $\tilde{a}=0$ (solid), 
	$0.5$ (dashed), $0.8$ (dot-dashed) and $0.998$ (dotted). Since
	the latter value is greater than $\tilde{a} >
	2 \pqty{\sqrt{2}-1}$, the spike
	extends into the ergosphere (shaded region).}
  \label{fig:rho_eqorbs}
\end{figure}

\begin{figure}[h]
	\begin{center}
		\input{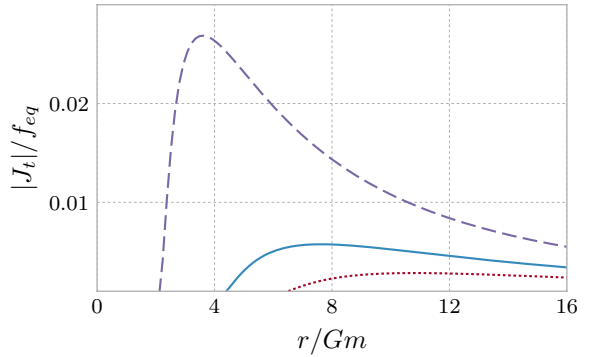}
	\end{center}
	\caption{The corotating (dashed) and counter-rotating (dotted) parts 
	of $\vqty{J_t}$ for $\tilde{a}=0.8$, compared with $\vqty{J_t}/2$ for
	the Schwarzschild case (solid).}
  \label{fig:jo_plus_minus}
\end{figure}

As further discussed in appendix~\ref{ap:equatorial}, 
this is a useful simplification 
because the numerical difficulty of the calculation is transferred to the 
pair of functions $\lcrit{}^\pm(\en)$.
Results for the density and the co- and counter-rotating parts of $J_t$ are 
shown in Figs.~\ref{fig:rho_eqorbs} and~\ref{fig:jo_plus_minus}. 
Figure~\ref{fig:rho_eqorbs} shows that increasing spin increases the density, 
and \fig{jo_plus_minus} is helpful in understanding the physical 
origin of this effect.

\begin{figure}[h]
	\begin{center}
		\input{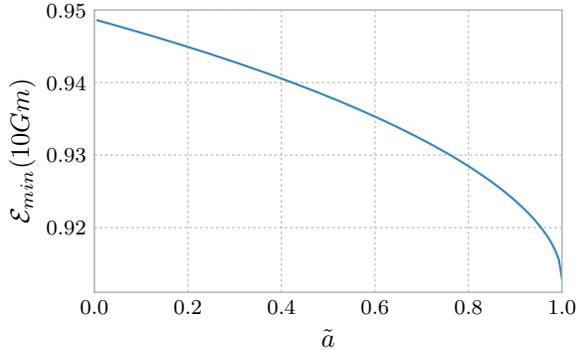}
	\end{center}
  \caption{Minimum allowed energy for $r/Gm=10$ as a function of Kerr parameter
	$\tilde{a}$ in the equatorial plane.}
  \label{fig:binding}
\end{figure}

As \fig{jo_plus_minus} shows, the density enhancement is coming 
from the corotating orbits, which are more deeply bound to the black hole. 
Since this binding energy increases with the spin parameter, the density will 
also increase with black hole spin. This is illustrated in 
\fig{binding}, which shows the minimum allowed energy for 
$r/Gm=10$ as a function of spin parameter $\tilde{a}$.

As explained in SFW, the density is zero at the coordinate $r$ such that 
$\emin(r)=1$. Orbits that go any closer to the black hole will have 
to be unbound in order to not be captured, and are therefore not included in 
this calculation. For the equatorial plane, this occurs at 
$r_{\text{mb}}^\pm/Gm=2\mp \tilde{a}
+2\sqrt{1\mp \tilde{a}}$, which corresponds to the radius of the marginally 
bound circular orbit, where the upper (lower) sign is for corotating
(counter-rotating) orbits. Note that the density vanishes at 
$r_{\text{mb}}^+$, 
which can be inside the ergosphere for $\tilde{a} \geq 2 \pqty{\sqrt{2}-1}
\approx 0.83$. Only corotating orbits contribute in the range 
$r_{\text{mb}}^+ \leq r \leq r_{\text{mb}}^-$. In a small region around 
$r\gtrsim r_{\text{mb}}^-$, when 
counter-rotating orbits start contributing, the slope of the density profile 
becomes less steep. For Schwarzschild, 
$r_{\text{mb}}^+= r_{\text{mb}}^- = 4 Gm$, which is where
SFW found the end point of the spike in this case.

\begin{figure}[h]
	\begin{center}
		\input{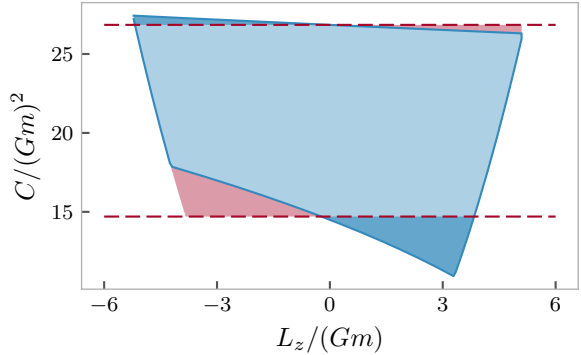}
	\end{center}
	\caption{The blue region shows a phase space slice of 
	fixed energy, $\en=0.98$, 
	at $r/Gm=20$ on the equatorial plane for a Kerr hole with 
	$\tilde{a} =0.5$. 
	The upper dashed line corresponds to $\vpot =0$ for 
	$r = 20 G m$, and the lower dashed line corresponds to the capture
	condition for $\en=0.98$, both in the Schwarzschild case.
	The red-shaded lower-left region shows the counter-rotating orbits
	that are lost due to capture by the hole, which are compensated by the
	tightly bound corotating orbits in the dark blue-shaded region.}
  \label{fig:phase_space}
\end{figure}

It is important to note that the boost obtained here is amplified by the fact 
that the corotating equatorial orbits are the most bound to the black hole. 
Nevertheless, when we integrate over the full phase space, the net effect is 
still the same: the enhanced binding of the corotating orbits is large enough 
to make up for the loss of counter-rotating orbits, which are preferentially 
captured by the black hole. This last point is also illustrated in 
\fig{phase_space}, which shows a fixed energy 
slice on the equatorial plane. The planar orbits fall on the left
and right
boundaries of the blue region, but the trade-off between loss of
counter-rotating orbits that is more than compensated by the addition
of corotating ones holds in general.

\subsection{Full phase space}
\label{sec:fullfconstant}

Unlike the equivalent calculation for the Schwarzschild case done in SFW, 
there is no analytic expression for the integration volume in the Kerr
geometry. None of the integrals can be performed analytically to 
simplify the current density. Therefore, we used a Monte Carlo (MC) method to 
overcome these difficulties. To maximize efficiency in the MC evaluation of 
the integrals, we look for the (\en, \carter, \lz)-cube that has the tightest 
fit to our phase space: we know that the maximum value of $\en$ is 1. We find 
an upper bound for \carter\ by noting that, from the positivity of 
\upot\ one obtains $\carter \geq \lz^2$ for bound orbits.

Thus, substituting $\lz=-\sqrt{\carter}$ and $\en=1$ in \vpot, one can obtain 
an upper bound on \vpot\ and, consequently, on \carter, as long as 
$r/Gm > 2$. For $r/Gm<2$ or, in general, for any $r$ within the ergoregion of 
the black hole, this upper bound on \carter\ can be found by looking for a 
plunge orbit with $\en=1$ and $\vpot=0$.

Once we have the upper bound on \carter\ and \en,
we find the orbit with minimum energy and Carter constant 
$\pqty{\emin(r,\theta),\cartercrit(r,\theta)}$ by requiring that all of the 
bounds in our phase space be satisfied simultaneously. This means 
$\vpot=\upot=0$, and this orbit must also be a plunge orbit, implying that 
the potential will have a double root at its unstable orbit $r_{\text{unst}}$.
This gives us a system of four polynomial equations for 
$\pqty{\emin,\cartercrit,\lz^*,r_{\text{unst}}}$, which we solve using a 
homotopy continuation method~\cite{Lee2008}.

We find it advantageous to implement the positivity of \upot\ explicitly
through the change of variables:
\begin{align}
  \en &= x+(1-x)\emin, \nonumber \\
	\carter &= y\carter_{\text{max}}+(1-y)\cartercrit, \nonumber \\
  L_z &= (2z-1)\sin\theta\sqrt{\carter-a^2\cos^2\theta(1-\en^2)},
  \label{eq:Duffy}
\end{align}
which puts our $(x,y,z)$-integration region in 
$[0,1]\times [0,1] \times [0,1]$. 

The equivalent for a general inclination of the radius of the marginally 
bound circular orbit is found by setting $\emin(r,\theta)=1$. The dark 
matter density vanishes for $r \leq r_{\text{min}}(\theta)$, since
no bound orbit exists within this region. 

Given a spin parameter $\tilde{a}$ and an inclination $\theta$, we can run 
the MC to calculate the current density for any $r>r_{\text{min}}(\theta)$ 
with these simplifications. 
We use two standard numerical routines, VEGAS~\cite{Lepage:1977sw} and 
MISER~\cite{Press:1989vk}, as implemented
in the GNU Scientific Library~\cite{gsl}. When evaluating the density at a
particular point, the numerical integrator performs $10^7$ function calls
to reach a relative accuracy at the level of $\lesssim 1\%$ across the entire 
$r$, $\theta$ range. 

\begin{figure}[h]
	\begin{center}
		\input{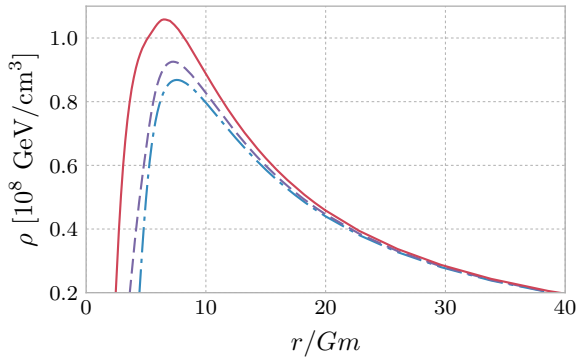}
	\end{center}
  \caption{The dark matter density in the equatorial plane increases with
	the spin parameter, and the spike gets closer to the hole. 
	The different lines show the Schwarzschild
	calculation from SFW (blue, dot-dashed), $\tilde{a}=0.5$ (purple,
	dashed),
	and $\tilde{a}=0.8$ (red, solid).}
  \label{fig:eq_constf}
\end{figure}

Sample results are displayed in Figs. 
\ref{fig:eq_constf} and \ref{fig:ani_constf}, where we use the same value
of $\flz=f_0=\frac{0.3\text{ GeV/cm}{}^3}{\pqty{2\pi (100 
\text{ km/s})^2}^{3/2}} = 5.1\times 10^8 \text{ GeV/cm}{}^3$
as in SFW to allow a direct comparison of our results. 
The density plot in \fig{disk_constf} provides a pictorial 
illustration of the density distribution.

Once more we find that the density increases with the spin parameter
$\tilde{a}$, and the spike gets closer to the hole.

\begin{figure}[h]
	\begin{center}
		\input{fig_rho_theta.pgf}
	\end{center}
	\caption{Density anisotropy for $\tilde{a}=0.8$. The spike is shown
	at different angles with respect to the black hole rotation axis:
	on axis $\theta=0$ (blue, dot-dashed), $\theta=\pi/3$ (purple, dashed)
	and equatorial $\theta=\pi/2$ (red, solid).}
  \label{fig:ani_constf}
\end{figure}

\begin{figure}[h]
	\begin{center}
		\input{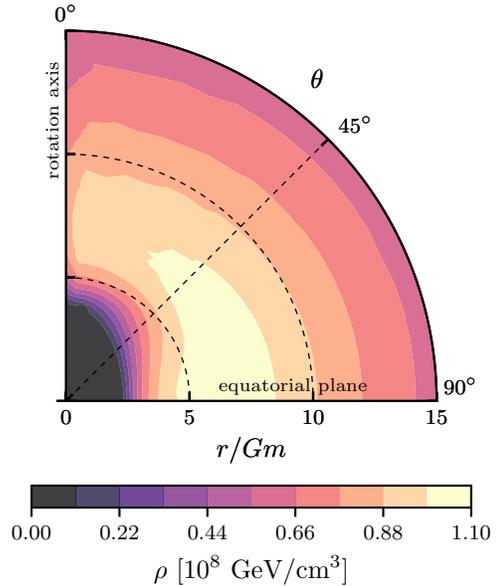}
	\end{center}
	\caption{Dark matter density in the $r-\theta$ plane for a spin
	parameter $\tilde{a}=0.8$, and a constant initial distribution
	function. The axis of the black hole points vertically, and
	$r/Gm$ is plotted from $0$ to $15$. The density is axisymmetric about
	the spin axis.}
  \label{fig:disk_constf}
\end{figure}

The decrease of the density as we get away from the equatorial plane is to be 
expected: for instance, only orbits with $\lz=0$ can cross the axis. Unlike 
the spherical case, in which the components $L_x$ and $L_y$ of the angular 
momentum are also conserved, this restriction effectively reduces the 
available phase space and, consequently, the density. This is useful since 
the calculation for orbits that cross the axis is simpler and provides us 
with a lower bound on the density everywhere.
Note that, as can be seen from comparing Figs.~\ref{fig:eq_constf}
and~\ref{fig:ani_constf}, although lower than on the plane, the density 
on the axes is still boosted for $\tilde{a}=0.8$ compared to the case of
Schwarzschild.

Indeed, using the last substitution in \eq{Duffy} and setting $\sin \theta=0$ 
allows us to perform the integrals over $z$ and 
\carter\ explicitly, obtaining:
\begin{align}
  \label{eq:axis}
  J_t=\frac{4\pi f_0}{\sqrt{\Delta}}
	\int\limits_{\emin}^1 \dd{\en} \en \sqrt{\carter_{max}(\en)-
	\cartercrit(\en)},
\end{align}
where, as usual, $\carter_{max}$ is obtained from $\vpot=0$ and
\cartercrit\ is the critical Carter constant. Because $\lz=0$, these are 
much simpler constraints than in the general case: \cartermax\ can be 
found algebraically and \cartercrit\ requires numerically solving a 
straightforward nonlinear equation. 
On the axis, $J_\phi=0$,
and the density is easily found from \eq{density} with $\Omega=0$. This
alternative route to calculate the density along the rotation axis of the
black hole provides us with a useful check of the full MC evaluation.

\section{Growth from a cuspy Hernquist profile}
\label{sec:hernquist}

The constant distribution is an adequate proxy for a cored 
profile~\cite{Gondolo:1999ef}, but DM-only simulations tend to 
favor cuspy distributions, and we would like to know if the calculated effect 
is still sizeable for a cuspy profile. Following SFW, we consider a Hernquist
profile~\cite{Hernquist:1990be}:
\begin{equation}
	\rho_{\text{H}} = \frac{\rho_0}{\pqty{r/r_s} \pqty{1+r/r_s}^3},
	\label{eq:rho_hernquist}
\end{equation}
which generates the Newtonian gravitational potential
\begin{equation}
	\Phi_{\text{H}} = -\frac{G M}{r_s+r}.
	\label{eq:pot_hernquist}
\end{equation}
Here $\rho_0$ and $r_s$ are scale factors related to the total
dark matter mass in the halo by $M = 2 \pi \rho_0 r_s^3$, which we take to
be $M=10^{12} M_\odot$ for the Milky Way.
The ergodic distribution function associated with the Hernquist 
profile can be found 
analytically~\cite{Hernquist:1990be}:
\begin{equation}
	\fhern =\frac{M}{\sqrt{2}(2\pi)^3(GMr_s)^{3/2}} \ftildehern,
	\label{eq:fhern}
\end{equation}
with
\begin{align}
	\ftildehern&=\frac{\sqrt{\tilde{\ehern}}}{(1-\ehern)^2}
	\bqty{ \pqty{1-2\ehern} \pqty{8\ehern^2-2\ehern-3} 
	+\frac{3\sin^{-1}\sqrt{\ehern}}{\sqrt{\ehern
	\pqty{1-\ehern}}} },
	\label{eq:ftildehern}
\end{align}
and we have introduced a new dimensionless relative energy 
$\ehern=-r_s E/GM$, which is related to the relativistic energy \en\ per
unit particle mass by 
\begin{equation}
	\ehern \equiv \frac{r_s}{GM} \pqty{1-\en}.
	\label{eq:enonrel}
\end{equation}
The halos found in simulations can be better fit by an NFW 
profile~\cite{Navarro:1995iw} with strongly correlated $\rho_0$, $r_s$.
As a result, halos are essentially members of a one-parameter family.
For a galactic mass halo, $r_s=20$ kpc, and $\rho_0$ is then fixed by the 
total mass. 
We choose to work with
a Hernquist profile with the same parameters because, as mentioned above, the 
distribution function can
be calculated analytically, and both halos have the same cuspy 
$\propto 1/r$ behavior in the inner region $r\lesssim r_s$ giving rise to the 
same Newtonian spike~\cite{Quinlan:1994ed}. 

Thus the only numerical difficulty introduced by this distribution is the 
evaluation of the radial action. 

Using \eq{pot_hernquist} in \eq{adb_invars_nr}, we can write the radial 
invariant for the Hernquist potential as:
\begin{equation}
	I_r^{\text{H}}=2\sqrt{GMr_s}\int\limits_{x_-}^{x_+} \dd{x}
	\left(\frac{2}{1+x}-2\ehern
	+\frac{\tilde{L}^2}{x^2}\right)^{1/2},
	\label{eq:irhern}
\end{equation}
where we introduced the dimensionless quantities $\tilde{L}=L\sqrt{GM r_s}$, 
$x=r/r_s$; and $x_{\pm}$ are the turning points of the orbit. As discussed
in \sect{adb},
$\tilde{L}$ and \ehern\ are obtained through the matching of
\eq{adb_invars_nr} to \eq{adb_invars} for each point in the Kerr phase 
space (\en, \carter, \lz) with the correspondence \eq{enonrel}. 

As $\en \to 1$, the orbit becomes unbound and the corresponding 
$\ehern \to 0$. In this limit, both radial invariants 
diverge as the upper turning point goes to infinity. To prevent numerical 
instabilities in the evaluation of the integrals, we follow SFW and
remap the interval $[x_-,x_+]$ to $[0,1]$ at each step of the bisection method
that is used to evaluate~\eq{irhern}. 
We do a similar remapping to evaluate the Kerr invariant\footnote{Note that 
the Schwarzschild limit of the radial invariant in Eq.~(3.19) in SFW 
is missing a factor of $r^2/\Delta = 1/(1-2 Gm/r)$.}.

The Hernquist distribution has most of its contribution to the density coming 
from the more deeply bound orbits, which are those that have $\ehern$ 
closer to 
1. Since we have argued in \sect{constf} that the presence of a 
deeper potential well for corotating orbits is what drives the enhancement of 
the spike, we expect it to increase more rapidly with increasing Kerr 
parameter. This is seen in Figs. \ref{fig:eqt_hern} and \ref{fig:ani_hern} 
below: the peak of the spike with $\tilde{a}=0.8$ is approximately 35\% 
higher than the one with $\tilde{a}=0.5$, whereas the corresponding boost for 
the constant distribution is around 20\%. Fig.~\ref{fig:figquarter_hern} shows
the density distribution resulting from an initial Hernquist profile.

\begin{figure}[h]
	\begin{center}
		\input{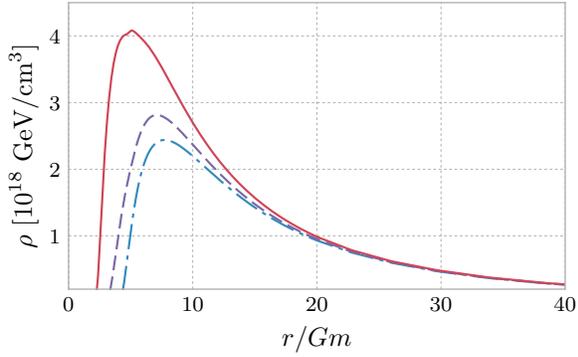}
	\end{center}
	\caption{Dark matter density in
	the equatorial plane with increasing Kerr parameter for an 
	initial Hernquist distribution. 
	The different lines show the Schwarzschild
	calculation from SFW (blue, dot-dashed), $\tilde{a}=0.5$ (purple,
	dashed),
	and $\tilde{a}=0.8$ (red, solid).}
  \label{fig:eqt_hern}
\end{figure}

\begin{figure}[h]
	\begin{center}
		\input{fig_angle_hern.pgf}
	\end{center}
	\caption{Density anisotropy for an initial Hernquist profile and
	$\tilde{a}=0.8$. The spike is shown
	at different angles with respect to the black hole rotation axis:
	on axis $\theta=0$ (blue, dot-dashed), $\theta=\pi/3$ (purple, dashed)
	and equatorial $\theta=\pi/2$ (red, solid).}
  \label{fig:ani_hern}
\end{figure}

\begin{figure}[h]
	\begin{center}
		\input{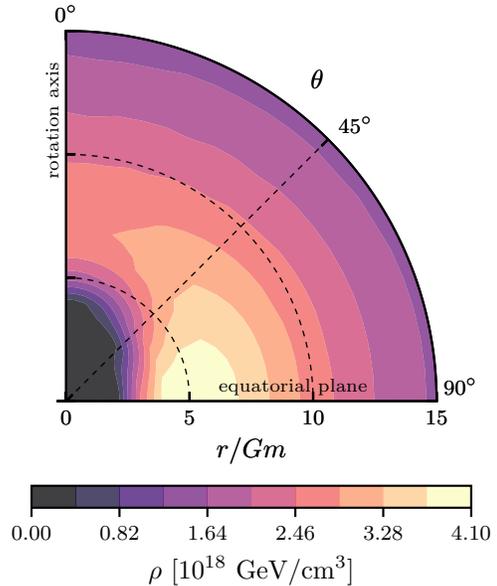}
	\end{center}
	\caption{Dark matter density in the $r-\theta$ plane for a spin
	parameter $\tilde{a}=0.8$, and an initial distribution
	function corresponding to a Hernquist profile.}
  \label{fig:figquarter_hern}
\end{figure}

\section{Discussion}
\label{sec:discussion}

The spatial extension of the 
spike is of the order of the radius of gravitational influence of the 
black hole, $r_h=Gm/\sigma_v^2 \sim 1.7$ pc for the galactic center black
hole, and the mass of the dark matter 
contained in this small volume is negligible compared to the mass 
of the black hole. Hence, as pointed out by SFW, the 
additional precession rate induced by the spike on the orbits of stars in the
central cluster will be small, and this conclusion remains valid for a 
rotating black hole. As shown in Fig.~5 in SFW, the rapid fall off with
distance of the effects of 
frame dragging and departure from spherical symmetry makes their
contribution to the pericenter advance of the orbits of stars subdominant 
when compared to 
the monopole Schwarzschild 
component over the range of semi-major axis $\gtrsim 0.1$ pc. The enhanced
dark matter density due to the rotation of the hole, will neither interfere 
with a test of the black hole no-hair theorem using
hypothetical stars with semimajor axes $\lesssim 0.2$ mpc~\cite{Will:2007pp},
nor change the conclusions of a future experiment sensitive to 
precession rates at the level of 10 $\mu$arcsec per year that could discover
the perturbing effects of the dark-matter for S2-type stars,
with semimajor axes $\sim 10$ pc~\cite{Sadeghian:2013laa}.

On the other hand, the enhanced density in the Kerr geometry 
could significantly alter the fluxes of high-energy radiation from dark matter
annihilations in the central regions surrounding the black hole. 
A prediction of the effects of rotation 
depends on the underlying particle physics model and on
the formation history of the supermassive black hole (see e.g.
\cite{Fornasa:2007nr} for a review).

As mentioned above, 
we are assuming that the growth of the black hole 
is adiabatic, but several dynamical effects could weaken the density
spike around the supermassive black hole.
Although these effects are important, our main purpose is to understand the
general relativistic effects close to the rotating black hole, and 
the following considerations provide an order of magnitude estimate in a 
sample of representative scenarios.

To estimate the effects of the spike on the dark matter annihilation fluxes
we compare the line of sight integral for a beam of opening angle $\theta$
towards the Galactic Center:
\begin{align}
	J(\theta)=\frac{\langle \sigma v \rangle}{4\pi m_{\chi}^2}
	\int\limits_0^{2\pi} \dd{\phi}\int\limits_0^{\theta}\dd{\psi}
	\cos\psi \sin\psi\int \dd{s} \rho^2 \pqty{r,\vartheta},
	\label{eq:jfactor}
\end{align}
with or without the presence of the spike. In \eq{jfactor},
$r=\sqrt{R^2+s^2-2Rs\cos\psi}$ is the Boyer-Lindquist coordinate, 
$s$ is the radial coordinate from the Earth 
to the annihilation point, $R=8.5$ kpc is the distance from the Earth to the 
Galactic Center, and $\vartheta$ is the angle relative to the equatorial 
plane of the black hole.

For a thermal relic,$\langle \sigma v \rangle = 3\times 10^{-26}cm^3/s$,
of mass $m_{\chi}=100GeV$,we find a flux of $6.3\times10^{-9}$
cm${}^{-2}$s${}^{-1}$ for an opening angle of $1^\circ$ from annihilations
in the halo with only the underlying Hernquist profile included.

To evaluate the integral in \eq{jfactor}, we fit the profiles that were
previously calculated using MC techniques to profiles of the form
\begin{align}
	\rho(r,\vartheta)=\frac{A}{x^p}
	\left(1-\frac{{r_{\text{min}}}(a,\vartheta)}{r}\right)^n,
	\label{eq:rhofit}
\end{align}
with all coefficients being allowed to vary with $\vartheta$. 
The expression in \eq{rhofit} is a generalization of Eq.~(9) 
in~\cite{Gondolo:1999ef} to allow for a $\theta$-dependent end point for
the spike. This expression is then matched to the power law 
$B/x^{\gamma_{\text{sp}}}$ using smooth functions to improve the fit and 
to give a reasonable estimate of $\rho$. We use 
$\gamma_{\text{sp}}=2.33$ corresponding to the Newtonian spike generated
by a $1/r$ NFW or Hernquist cusp.

We extend our spike profile until the density is equal to that of the
underlying Hernquist profile. This happens at $12.4$ pc, and we take the 
fiducial
$1/r$ halo shape beyond this point. Putting all the pieces together, our
model for the central part of the halo is equivalent to the canonical model
used in~\cite{Fields:2014pia}, aside from the fact that we are assuming
an underlying Hernquist profile with $r_s=20$ kpc and the extension
of the spike. For the Schwarzschild geometry, the spike enhances the flux
by a factor of $1.93\times10^9$ relative to the initial 
Hernquist profile. In Table~\ref{tb:boost} we present the
flux enhancement normalized to the Schwarzschild one, as that is less
sensitive to the normalization of the underlying density profile.

Note that, as mentioned above, the 
influence radius of the black hole is $r_h \sim 1.7$ pc, so at these 
distances the spike should start being modified. Also,
our calculation of the Kerr spike does not consistently take into account
the gravitational field generated by the dark matter distribution itself.
We leave for future work a proper treatment of this effect along the lines of
the grid calculation in~\cite{Quinlan:1994ed} for the nonrelativistic case.
Let us stress that it is only in the transition region between the black hole
dominated field and the smooth underlying halo that we expect significant
changes. Moreover, the relative boost factors quoted in Table~\ref{tb:boost}
will remain mostly unchanged.

Our expression for the spike does not include the effects of dark matter
annihilations which will deplete and weaken the density profile. SFW followed
the strategy used in~\cite{Gondolo:1999ef}, and considered a constant core
within the radius $r_{\text{ann}}$ determined by the location where the 
density equals $\rho_{\text{ann}} = m_{\chi}/\sigma v t_{\text{bh}}$, which
for our thermal relic turns out to be $r_{\text{ann}}=1.3\times 10^{-2}$ pc,
assuming that the annihilation process has been acting over $t_{\text{bh}}=
10^{10}$ yr. It was pointed out in~\cite{Vasiliev:2007vh} that a plateau is
in equilibrium if all the dark matter particles move in circular orbits,
and the density forms a "weak cusp" $\propto r^{-1/2}$ for a more realistic
isotropic distribution of DM velocities. This behavior has been confirmed
by integrating the Boltzmann equation in~\cite{Shapiro:2016ypb} for the case
of $s$-wave annihilation. If we assume that the DM is distributed as
$r^{-1/2}$ within $r_{\text{ann}}$ then the density is sufficiently low that
the spike goes away. The effects of the black hole spin are therefore washed
out and we get the same boost factor for a Schwarzschild as for a nearly
extreme $\tilde{a}=0.998$ black hole. In both cases the profile close to the
hole is the same $r^{-1/2}$ weak spike, and the ratio of the flux coming from
the
spike to that coming from the smooth halo is $2.8\times 10^3$. This large
enhancement is due to the spatial extension of our spike, which shows that
most of the annihilation signal comes from the outer regions of the spike, where
relativistic corrections are not important.

The weak $r^{-1/2}$ is not generic, but depends on the $s$-wave nature of the
annihilating process. For $p$-wave annihilation, the final cusp is even weaker
and again erases any effects due to the Kerr spin. On the other hand, a
possible detection of an identified x-ray line at $E \approx 3.55$ keV from
the Perseus cluster and the Andromeda 
galaxy~\cite{Bulbul:2014sua,Boyarsky:2014jta} have been attributed to a
sterile neutrino~\cite{Dodelson:1993je}. This dark matter candidate decays 
emitting an x-ray line and its density profile is the Kerr spike without
any attenuation due to self-annihilation. The x-ray flux due to dark matter
decays, however, is proportional to the density as opposed to the 
density squared behavior of the annihilation signal shown in \eq{jfactor}.
Much like in the case of the precession rates induced by the spike on the
orbits of stars, we expect the additional decay flux induced by the spike 
to be small, and the effects due to the rotation of the black hole to be
subdominant.

\begin{table}
	\begin{tabular}{|c | c |}
		\hline
		$\tilde{a}$ & $J(a)/J(0)$ \\
		\hline
		$0$ & $1$\\
                $0.5$ & $1.11$\\
		$0.6$ & $1.14$\\
		$0.7$ & $1.22$\\
		$0.8$ & $1.38$\\
		$0.9$ & $1.59$\\
		$0.998$ & $1.97$\\
		\hline
	\end{tabular}
	\caption{Boost factors for different Kerr spin parameters $\tilde{a}$
	for the full spike with no annihilation normalized to the 
	Schwarzschild spike.}
	\label{tb:boost}
\end{table}

Self-interacting dark matter~\cite{Spergel:1999mh} with a cross section as 
large as $\sigma/m_{\chi}\approx 0.1$ cm${}^2$/g has been invoked to address
several discrepancies between numerical predictions of cold dark matter 
models and observations of subgalactic scale 
structures~\cite{Feng:2009mn,Rocha:2012jg}. 
The effects of self-interactions have been shown in~\cite{Shapiro:2014oha} to
replenish the weak cusp, giving rise to a steeper profile depending on
the velocity dependence of the cross section.
Since a detailed study of the final relaxed distribution using the exact 
phase-space distribution of the Kerr spike is missing,
we display in Table~\ref{tb:boost} the radiation fluxes as arising from a 
full Kerr spike with no weakening due to annihilation. The results 
show that when going from a nonrotating to an almost extremal black hole,
the boost factor is almost doubled. 

\section{Conclusions}
\label{sec:conclusions}

We have extended the analysis performed in SFW to include the effects of 
black hole spin. Our findings show that the spike persists around a rotating 
black hole and, furthermore, that it is enhanced. 

\begin{figure}[H]
	\begin{center}
		\input{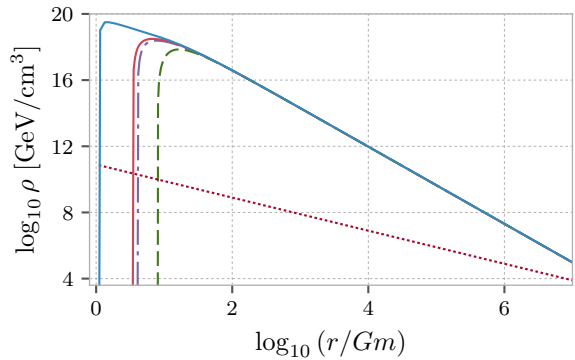}
	\end{center}
	
	\caption{Equatorial (solid blue) and on-axis (solid red) density 
	obtained for the near-extreme black hole, compared with previous 
	calculations assuming a Schwarzschild hole (dot-dashed purple), and
	the nonrelativistic estimate in~\cite{Gondolo:1999ef} (dashed green).
	The initial DM distribution before the growth of the black hole is
	a Hernquist profile shown as the dotted red line.}
  \label{fig:summary}
\end{figure}

Since the total mass contained in the spike is not very large, effects that 
depend on the total mass of the spike such as the stellar precession studied 
in \cite{Sadeghian:2013laa} or fluxes from decaying dark matter will 
essentially remain unaltered by the inclusion of rotation. Our results are 
summarized in \fig{summary}, which shows the largest density 
enhancement, obtained for a near-extreme black hole ($\tilde{a}=0.998$) in the 
equatorial plane, as well as the density along the spin axis, comparing them 
to previous calculations.

However, the large growth of the spike could have consequences for 
observables related to dark matter annihilation, which depend on the density 
squared. Further work remains to be done here to properly implement the 
evolution of the spike and the presence of a weak 
cusp \cite{Shapiro:2016ypb,Vasiliev:2007vh}, which are both necessary for the 
extraction of a gamma-ray signal.

Although we have focused our attention on the signals from
the Galactic Center, black holes are ubiquitous in nature. Our findings could
potentially impact the cumulative effects of dark matter spikes on the diffuse
gamma-ray background~\cite{Belikov:2013nca}, or the signals from local dwarf
spheroidal galaxies~\cite{Gonzalez-Morales:2014eaa}.

\acknowledgments

We are grateful to Tongbo Liu, Oleg Ruchayskiy, Laleh Sadeghian, Lijing Shao
and Jessie Shelton
for useful discussions.
F. F. acknowledges the hospitality of the Aspen Center for 
Physics, which is supported by the National Science Foundation.
F. F. and A. M. R. were supported in part by the U.S. Department of Energy, 
Office of
High Energy Physics, under Award No. 
DE-FG02-91ER40628. C. M. W. was supported in part by the National Science
Foundation under Grant No. 16-00188.

\appendix

\section{Adiabatic invariance of the distribution function}
\label{ap:adiabatic_f}

We here prove the important property that the distribution function is
an adiabatic invariant, which we use in our code.
A proof for nonrelativistic 
spherical systems is given in Young~\cite{Young:1980apj}, which was
generalized to the general relativistic Schwarzschild case by 
Sadeghian~\cite{Sadeghian:2013bga}. The extension to the Kerr geometry uses 
the same ideas, but the change of variables that must be performed in the 
adiabatic evolution is two-dimensional instead of one-dimensional.

We start by integrating the current density \eq{final_current} to find the
mass enclosed in a hypersurface of constant time. The future-pointing normal
vector to the surface is $n_{\alpha}=-(-g^{tt})^{-1/2}\partial_{\alpha}t$, 
and the three-dimensional surface element can be written 
as~\cite{2009PhyA..388.1818D}:
\begin{equation}
	\label{eq:tconstant}
	\dd[3]{S_\alpha} = -\delta^0_\alpha \frac{\sqrt{g_S}}{\sqrt{-g^{tt}}}
	 \dd[3]{x},
\end{equation}
where 
\begin{equation}
	g_S = \frac{\Sigma^4}{\Delta} 
	g_{\phi\phi},
\end{equation}
is the determinant of the metric induced on the hypersurface.

The enclosed mass is therefore:
\begin{align}
	\label{eq:mass}
	M &=-\int \dd[3]{S_\alpha} J^{\alpha} \notag \\
	&= \int \dd{r} \dd{\theta} \dd{\phi} \frac{\sqrt{g_S}}{\sqrt{-g^{tt}}}
	\left(g^{tt} J_0 + g^{t\phi} J_\phi\right) \notag\\
	& = 2\int \frac{\dd{r} \dd{\theta} \dd{\phi} \dd{\en} \dd{\carter}
	\dd{\lz}}{r^2 \sin \theta \sqrt{\vpot} \sqrt{\upot}} 
	\frac{\Sigma^2\sqrt{g_{\phi\phi}}}{\sqrt{-\Delta g^{tt}}} \notag \\
	& \qquad \times \pqty{g^{t\phi}\lz-g^{tt}\en} \flz.
\end{align}

Interchanging the order of integration, the integrals over coordinates give us
a distribution $N(\en,\carter,\lz)$ of particles per unit conserved quantity. 
Using the relation $g^{tt}=-g_{\phi\phi}/(\Delta \sin^2\theta)$ for Kerr, we
obtain:
\begin{align}
	\label{eq:distribution}
	N(\en,\carter,\lz)&=4\pi\int \frac{\dd{r} \dd{\theta}}{r^2\sqrt{\upot}
	\sqrt{\vpot}}\Sigma^2 \notag \\
	& \qquad \times \pqty{g^{t\phi}\lz-g^{tt}\en} \flz.
\end{align}
In the above expression, the limits of integration are the radial turning 
points of \vpot\ and the angular turning points of \upot. Under adiabatic 
evolution, the constants of the motion will change to new values 
\en${}^*$, \carter${}^*$, \lz${}^*$, in such a way that 
$N\pqty{\en,\carter,\lz} \dd{\en} \dd{\carter} \dd{\lz} = 
N^*\pqty{\en{}^*,\carter{}^*,\lz{}^*} \dd{\en{}^*} 
\dd{\carter{}^*} \dd{\lz{}^*}$. The new values of the constants of the motion 
are determined by the invariance of the action integrals in \eq{adb_invars}. 
If we assume that the initial and final states are axisymmetric, then 
the invariance of $I_{\phi}$ gives $\lz=\lz^*$. 
The remaining two integrals 
allow us to
relate $N$ and $N^*$ through the chain rule of multivariable calculus, 
\begin{displaymath}
\frac{\partial (\en^*,\carter^*)}{\partial(\en,\carter)}=
	\frac{\partial(\en^*,\carter^*)}{\partial(I_r^*,I_{\theta}^*)}
	\frac{\partial(I_r,I_{\theta})}{\partial{(\en,\carter)}}.
\end{displaymath}
Taking partial derivatives in \eq{adb_invars}, we find:
\begin{align}
  \label{eq:Jacobian}
	\mathfrak{J}& \equiv \frac{\partial \pqty{I_r,I_{\theta}}}{
		\partial\pqty{\en,\carter}} \notag \\
	&=2\int \frac{\dd{r} \dd{\theta} \; 
	\Sigma^2 \pqty{\en g_{\phi\phi}+\lz g_{t\phi}}}{r^2 \sin^2\theta\Delta
	\sqrt{\upot}\sqrt{\vpot}}.
\end{align}
Using $g^{tt}=-g_{\phi\phi}/(\Delta\sin^2\theta)$, 
$g^{t\phi}=g_{t\phi}/(\Delta\sin^2\theta)$ and comparing to \eq{distribution}, 
we immediately obtain
\begin{displaymath}
	N \pqty{\en,\carter,\lz} = 2 \pi \mathfrak{J}\: \flz.
\end{displaymath}
Therefore, we find
\begin{align*}
  \label{eq:qed}
	N\pqty{\en,\carter,\lz}& \dd{\en} \dd{\carter} \dd{\lz}\\ 
	&= 2\pi\mathfrak{J}\: \flz \dd{\en} \dd{\carter} \dd{\lz} \\
	&= 2\pi\mathfrak{J}^* \: f^*\pqty{\en^*,\carter^*,\lz^*}
	\dd{\en{}^*} \dd{\carter{}^*} \dd{\lz{}^*}\\
        &=2\pi\mathfrak{J^*} \: f^*\pqty{\en^*,\carter^*,\lz^*} \\
	&\qquad\times
	\frac{\partial(\en^*,\carter^*)}{\partial(I_r^*,I_{\theta}^*)}
	\frac{\partial(I_r,I_{\theta})}{\partial{(\en,\carter)}}
	\dd{\en} \dd{\carter} \dd{\lz}\\
        &=2\pi\mathfrak{J^*} \: f^*\pqty{\en^*,\carter^*,\lz^*}
	\frac{\mathfrak{J}}{\mathfrak{J^*}}
	\dd{\en} \dd{\carter} \dd{\lz}\\
  	&=2\pi\mathfrak{J} \: f^*\pqty{\en^*,\carter^*,\lz^*}
	\dd{\en} \dd{\carter} \dd{\lz},
\end{align*}
which demonstrates the adiabatic invariance of the distribution function,
\begin{displaymath}
	f\pqty{\en,\carter,\lz}=f^*\pqty{\en^*,\carter^*,\lz^*}.
\end{displaymath}

\section{Phase space for orbits that remain on the equatorial plane}
\label{ap:equatorial}

We will set $G=1$ and work with the dimensionless quantities:
\begin{displaymath}
	\tilde{a}\equiv \frac{a}{m}\qquad \tilde{L}_z \equiv \frac{\lz}{m}
	\qquad \tilde{\carter}\equiv \frac{\carter}{m^2} \qquad x \equiv
	\frac{r}{m},
\end{displaymath}
and we will drop the tildes for the rest of this appendix.

The black hole horizon is located at 
\begin{equation}
	x_{\text{horizon}}= 1 + \sqrt{1-a^2},
	\label{eq:horizon}
\end{equation}
and the boundary of the ergosphere is
\begin{equation}
	x_{\text{ergosphere}} = 1 + \sqrt{1-a^2 \cos^2\theta},
	\label{eq:ergosphere}
\end{equation}
which is equal to $x_{\text{ergosphere}}=2$ for the equatorial 
$\theta = \pi/2$ latitude. 

For planar trajectories $\lz = \pm \sqrt{\carter}$, with the plus (minus) sign
corresponding to corotating (counter-rotating) orbits. 
Hence, the subset of orbits that we are considering is defined by
two quantities \en\ and \lz. These are constrained to satisfy $\en \leq 1$,
which bounds the energy from above;
$\vpot \geq 0$, which results in an upper bound on the angular momentum
\lmax; and the capture condition 
\begin{equation}
	V=0 = \dv{V}{x},
	\label{eq:capture}
\end{equation}
which will determine the critical angular momentum, \lcrit, and the minimum
energy, \emin. Note that the constraint $\upot \geq 0$ is trivially satisfied.

\begin{widetext}
The effective potential for orbits in the equatorial plane reads:
\begin{equation}
	\eval{V}_{\carter=\lz^2} = \frac{\en^2 \pqty{x^3 + a^2 (2 +x)}-
	4 a \lz \en + (2 -x) \pqty{\lz^2+x^2} -a^2 x}{x^3}.
	\label{eq:veff}
\end{equation}
We will use this expression to obtain \lmax, \lcrit\ and \emin, at a fixed
distance $x$ from the hole.
\end{widetext}

\subsection{$\lz^{\text{max}}$}

For a fixed energy $\en < 1$, the constraint $V \geq 0$ puts an upper bound
on \lz. 
We can explicitly find the boundary $V=0$,
since $V$ is quadratic in \en\ and \lz:
\begin{equation}
	\lz^{\text{max}} = \frac{- 2 a \en \pm \sqrt{ x \pqty{a^2-x(x-2)} 
	\pqty{2 - \pqty{1-\en^2} x}}}{x-2}.
	\label{eq:lmax}
\end{equation}
We generally obtain two solutions: a positive value of \lz, which is the 
maximum for corotating orbits, and a negative one
corresponding to the smallest \lz\ for counter-rotating orbits (which we also
name \lmax). But these solutions might not exist for all $x$, since the 
polynomial inside the square root can become negative. 
Being a quartic polynomial in $x$ with negative quartic coefficient, $\en^2-1$,
it will become negative (i.e. no $\lz^{\text{max}}$) at large $x$. This 
reflects the fact that far away orbits are closer to being unbound as
discussed below [c.f. \eq{ebound}].

More specifically, the discriminant in \eq{lmax} vanishes at the nonphysical 
points $x=0$ and $x = 1-\sqrt{1-a^2}$, which are inside the horizon. It also 
vanishes at
\begin{equation}
	x = x_{\text{horizon}} \quad \& \quad x = \frac{2}{1-\en^2} > 2,
	\label{eq:discroots}
\end{equation}
so the square root is well defined between these two locations and we will
find two values of $\lz^{\text{max}}$.

This is true for points outside of the ergosphere, $x > 2$. 
At the ergosphere, \eq{veff} is linear in \lz, and there is only one solution. 
A closer look shows that the positive root smoothly tends to this 
single root,
\begin{displaymath}
	\lmax=\frac{2 a^2 \en^2 + 4 \en^2-a^2}{2 a \en},
\end{displaymath}
which may be positive.

The negative branch in \eq{lmax} diverges as $x\rightarrow 2$.
This does not mean that counter-rotating orbits with arbitrarily large 
$\left| \lz \right|$ are allowed, since we will see that they cease
to exist that close to the hole.
For the Schwarzschild case SFW found 
that the critical value of the angular momentum limited orbits to lie beyond 
the unstable marginally bound orbit at $x=4$ for
$\en=1$. The minimum energy $\en=\sqrt{8/9}$ is attained at the location of
the ISCO, $x=6$. For
a Kerr hole the locations of these orbits get closer to the horizon for
corotating orbits ($x=1$ for the extremal $a=1$ case) and pushed away for 
counter-rotating ones ($x=5.83$ and $x=9$).
As we prove below, there are no orbits within the radius
of the marginally bound orbit,
\begin{equation}
	x_{\text{mb}} = 2 \mp a + 2 \sqrt{1\mp a},
	\label{eq:mborbit}
\end{equation}
and we will not have to worry about regions where \eq{lmax} does not provide 
a valid bound. Nevertheless, there might be corotating bound orbits within
the ergosphere since \eq{mborbit} can be $<2$.

Note also that, for a given $x$, there is a minimum value of the energy since,
for $V$ to be positive, we need $x$ to be to the left of the second root 
in \eq{discroots}, so 
\begin{equation}
	\label{eq:ebound}
	\en^2 \geq 1-\frac{2}{x}.
\end{equation}

\subsection{\lcrit\ and \emin}

Let us examine the capture condition, which is going to provide 
us with \lcrit\ and \emin.
The first equality in \eq{capture} is a cubic equation in $x$, while 
the second requires solving a 2nd order polynomial.

\begin{widetext}
The solution to the second equation is
\begin{equation}
	\dv{V}{x}=0 \Rightarrow x_\mp = \pqty{a^2 \pqty{1-\en^2} +\lz^2 \mp
	\sqrt{\pqty{\lz^2+a^2 \pqty{1-\en^2}}^2 - 12 (\lz -a \en)^2}}/2.
	\label{eq:xturn}
\end{equation}

The smaller root $x_-$ will correspond to the turning point when 
the constraint $V=0$ is also satisfied. It has the correct Schwarzschild limit:
\begin{displaymath}
	\lim_{x\rightarrow a} x_- = \frac{\lz^2 - \lz \sqrt{\lz^2-12}}{2} =
	\frac{6}{1+\sqrt{1-12/\lz^2}},
\end{displaymath}
which is the location of the unstable circular orbit.

We can now plug this value back in the equation $V=0$ and solve for \lcrit:
\begin{align}
	V& \pqty{x_,\en,\lcrit} = 0 \Rightarrow  \notag \\
	&\qquad 0=-9 (\lcrit-a\en)^2 \times \notag \\
	&\qquad\left[\pqty{\en^2-1} \lcrit{}^6 
	+ \pqty{36 \en^2-27 \en^4 -3 a^2
	\pqty{1-\en^2}^2-8} \lcrit{}^4\right.  
	+ 36 a \en \pqty{2-5\en^2+3 \en^4}
	\lcrit{}^3 \notag\\ 
	&\qquad +\pqty{2 a^2 \pqty{10-91 \en^2+162\en^4-81 \en^6} + 3 a^4
	\pqty{\en^2-1}^3-16} \lcrit{}^2 
	+ 4 a \en \pqty{8+9a^2 \pqty{1-\en^2}^2
	\pqty{-1+ 3\en^2}} \lcrit\notag\\ 
	&\qquad \left. - 16 a^2 \en^2 -a^6 \pqty{1-\en^2}^4
	-a^4 \pqty{1-\en^2}^2 \pqty{-1-18\en^2+27\en^4}\right],
	\label{eq:lcrit}
\end{align}
which contains a sixth-order polynomial in \lcrit\ that can be numerically
solved for given $a$ and \en. Since $\lcrit \leq \lmax$, we select the
largest of all the roots that are smaller than \lmax.
\end{widetext}

The value of \lcrit\ will depend on our choice of \en. For corotating orbits
it can be shown from \eq{lmax} that  \lmax\ is a monotonically 
increasing function of \en, so it will be largest for $\en=1$,
while \lcrit\ is a monotonically decreasing function of \en. On the other
hand, keeping the value of \lz\ constant, there is a minimum energy
\emin, which satisfies $V=0$.

\begin{figure*}[htb]
	\begin{center}
		\input{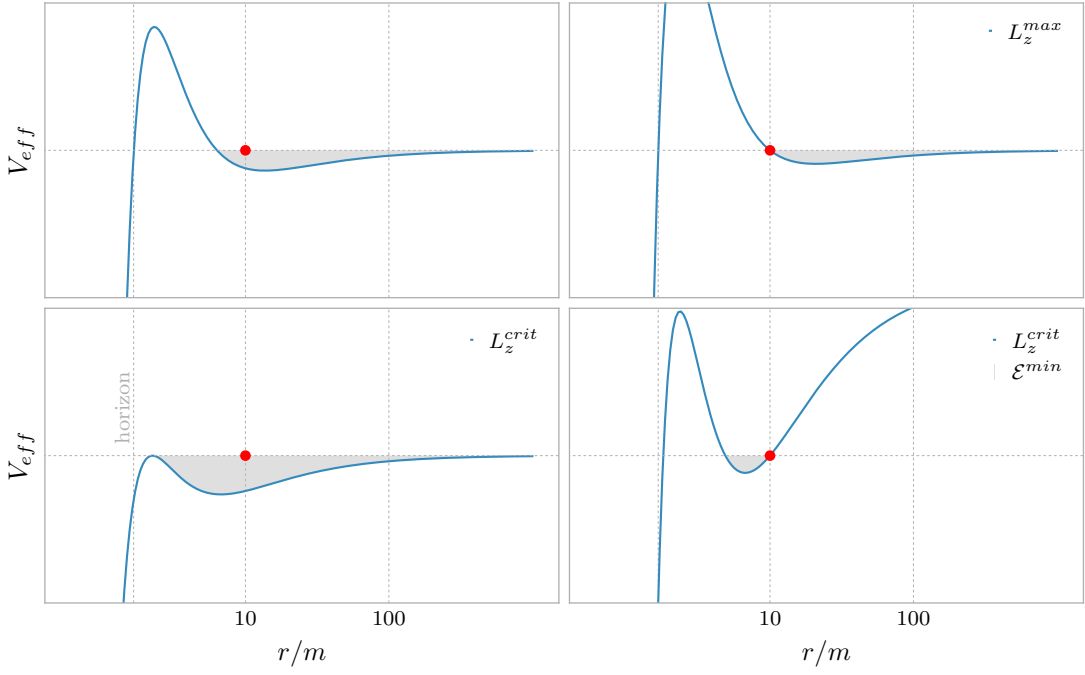}
	\end{center}
	\caption{Momentum-space at the location $x=10$, which is shown as a 
	red dot. {\em Top left:} 
	Starting from allowed values of $\en = \en_0 = 0.9997$ and $\lz=4$, 
	the red
	dot is well within the allowed region. {\em Top right:} For this
	value of $\en$, changing $\lz \rightarrow \lmax=4.829$ from
	\eq{lmax} pulls the effective potential $V_{\text{eff}}\equiv -\vpot$
	up and brings the red dot to the edge of the allowed region.
	{\em Bottom left:} Keeping $\en=\en_0$ but changing $\lz \rightarrow
	\lcrit=2.9$ brings the effective potential 
	down, so that the orbit is almost 
	captured. {\em Bottom right:} Once we have the $\lz$ bounds, 
	lowering the energy while keeping $\lz = \lcrit$ will pull the 
	effective potential curve up. When it
	touches the red dot we have found $\emin=0.938$. 
	Setting $\en_0=\emin$ the process can be iterated to find 
	$\lmax = 2.99$, $\lcrit=2.73$, and the next $\emin=0.931$.}
	\label{fig:viterate}
\end{figure*}

This allows us to find the boundary of momentum-space at the location $x$
using the following numerical scheme depicted in \fig{viterate}. 
Starting with $\en=1$,
\begin{enumerate}[i)]
	\item Find \lmax\ from \eq{lmax}, and \lcrit\ solving \eq{lcrit}. 
		As long as $\lmax \geq \lcrit$, the volume in momentum space
		at $x$ is nonzero.
	\item Solve $V\pqty{\lz = \lcrit}=0$ to find \emin.
\end{enumerate}
Note that we have found the minimum energy of all the orbits with $\lz=\lcrit$,
where \lcrit\ was found by setting $\en=1$. But we can now set $\en=\emin$
and repeat the steps above to find a smaller \lmax\ and a larger \lcrit;
and a smaller \emin. This process can be iterated until we get to an energy 
for which $\lmax = \lcrit$, this is the minimum energy for any \lz\ 
allowed at the location $x$.
An analogous strategy can be followed for counter-rotating orbits.

As we get closer to the hole, the phase space gets reduced. Orbits with
smaller values of \en, which were allowed at larger distances, are now pulled
in and trapped. Eventually, we reach a point where only a single
orbit with $\en=1$ is allowed. In this case, the iteration process above will
end immediately, since $\lmax =\lcrit$ for $\en=1$ at that point. We can 
find this location by solving for $x$ from the equation $\lmax=\lcrit$ with
$\en=1$. 

It turns out that this can be done analytically, since for $\en=1$
\eq{lcrit} simplifies to
\begin{displaymath}
	0=-9 \pqty{a -\lz}^2 \pqty{-16 a^2 + 32 a \lz - 16 \lz^2+\lz^4}.
\end{displaymath}
The roots of this polynomial can be found in closed form:
\begin{align}
	\lcrit &= a, 2 \pqty{1-\sqrt{1-a}},2 \pqty{1+\sqrt{1-a}},\notag\\
	&\quad 2 \pqty{-1-\sqrt{1+a}},2 \pqty{-1+\sqrt{1+a}}.
	\label{eq:lcritroots}
\end{align}
On the other hand, \eq{lmax} becomes for $\en=1$:
\begin{equation}
	\lmax = \frac{-2a \pm \sqrt{2}\sqrt{x \pqty{a^2 + (x-2) x}}}{x-2}.
	\label{eq:lmaxe1}
\end{equation}

We can solve for the values $x$ where $\lcrit=\lmax$ by
equating each of the roots in \eq{lcritroots} to each of the two branches in
\eq{lmaxe1}.
Equating the negative branch of \eq{lmaxe1} to the
only negative root in \eq{lcritroots} and solving for $x$ we find 
for counter-rotating orbits,
\begin{displaymath}
	x_{\text{min}}^{\lz<0}= 2+a+2\sqrt{1+a}.
\end{displaymath}

For corotating orbits, we equate the four positive roots in 
\eq{lcritroots} to the positive branch of \eq{lmaxe1}. The solution for
$\lmax=a$ occurs at $x=0,a^2/2$, which are inside the horizon and thus
nonphysical; $\lmax=2\pqty{\pm 1 \mp \sqrt{1 \mp a}}$ are located at
$x = 2 \pm a -2 \sqrt{1\pm a}$, which are also inside the horizon. The only
physical solution is 
\begin{displaymath}
	x_{\text{min}}^{\lz>0}= 2-a+2\sqrt{1-a}.
\end{displaymath}
The locations above coincide with \eq{mborbit}. Hence, the volume of
momentum-space available vanishes at the location of the marginally bound 
orbit, and only corotating orbits exist inside the ergosphere.

\begin{figure*}[htb]
	\begin{center}
		\input{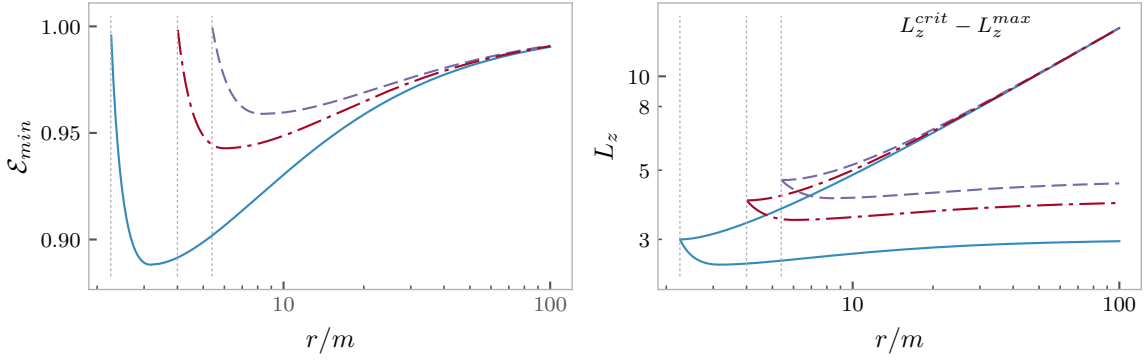}
	\end{center}
	\caption{Comparison of the edges of the phase-space volume as a
	function of distance $x=r/m$ from the hole for corotating orbits
	(solid blue), counter-rotating orbits (dashed magenta), and 
	Schwarzschild (dot-dashed red). The left panel shows \emin, while the
	range of allowed values of \lz\ is shown in the right one as
	the regions bounded by \lmax\ and \lcrit. The
	Kerr hole has spin $a=0.75$, and the sign of the angular momentum
	for the counter-rotating orbits has been flipped. The dotted vertical 
	lines mark the location of the marginally bound orbit, where the 
	phase-space vanishes. }
	\label{fig:phasespacecomp}
\end{figure*}

The boundaries derived above are shown in \fig{phasespacecomp}, which
also shows the Schwarzschild limit for comparison. Note 
that in the latter case, the absolute minimum energy 
$\en_{\text{min}}=\sqrt{8/9} = 0.94$ is attained at $x=6$ in agreement with 
the results in SFW.

\subsection{Density on the equatorial plane}

Let us consider the two-dimensional equatorial plane described by the
restriction of the Kerr geometry to $\theta=\pi/2$. The definition of
$J^\mu$ in \eq{define_J} is still valid, but now the indices run over
$\mu=t,r,\phi$ only. Our geodesics are described by the constants of motion
\en, \lz, $\mu$, and the Jacobian of the change is
\begin{equation}
  \label{eq:jacobian2d}
  \mathcal{J}=\left\rvert \frac{\partial(\en,\lz,\mu)}{\partial(p^t,
	p^r,p^{\phi})}\right\rvert
	=\frac{\Delta |u_r| \sin^2\theta}{\mu^2},
\end{equation}
where we are temporarily using dimensionful quantities to parallel the
discussion in the main text.
Now there is only an additional factor of $2$ to account for the sign in
$p^{\phi}$, and $\sqrt{-g} = r$. Putting things together and using
$\vqty{u_r} = r^2/\Delta \sqrt{\vpot}$, we obtain
\begin{displaymath}
	\sqrt{-g} \dd{p^t} \dd{p^r} \dd{p^\phi} = \frac{2 \mu^2}{r\sqrt{\vpot}}
	\dd{\en} \dd{\lz} \dd{\mu}.
\end{displaymath}
We have the relation $f^{(3)}(x,p) = \mu^{-2} f\pqty{\en,\lz} 
\delta \pqty{\mu - \mu_0}$, where $f\pqty{\en,\lz}$ has units
of surface mass density. Considering a constant distribution on the plane 
$f\pqty{\en,\lz}=\fplane$, we obtain
\begin{equation}
	\label{eq:generalcurrent2d}
	J_\mu = \int{f^{(3)}(x,p) u_\mu \sqrt{-g} \dd[3] p} = 
	\frac{2 \fplane}{r} \int{u_\mu \frac{\dd{\en} \dd{\lz}}{\sqrt{\vpot}}}.
\end{equation}

Going back to using dimensionless quantities, e.g. $\lz = G m
\tilde{\lz}$ and dropping the tildes again, we obtain more explicitly
\begin{align}
	\label{eq:current2d}
	J_t & = - 2 \fplane \sqrt{x} \int_{\emin}^1 {\en \dd{\en} 
	\mathfrak{I}_t (\en)},\notag \\
	J_\phi & =  2 \fplane \sqrt{x} (G m) \int_{\emin}^1 { \dd{\en} 
	\mathfrak{I}_\phi (\en)},
\end{align}
where
\begin{align}
	\label{eq:lzintegral}
	\mathfrak{I}_t (\en) &\equiv 
	\int_{\Delta \lz} \frac{\dd{\lz}}{\sqrt{\vx}} \notag \\
	\mathfrak{I}_\phi (\en) &\equiv 
	\int_{\Delta \lz} \frac{\lz \dd{\lz}}{\sqrt{\vx}},
\end{align}
and \vx\ is the numerator in \eq{veff}. The \en-dependent
region of integration
$\Delta \lz = \pqty{\lminus,\lcritm} \cup \pqty{\lcritp,\lplus}$ 
includes both corotating and counter-rotating orbits, and \lplus, \lminus\ 
are the respective values of \lmax\ in \eq{lmax}.

\begin{widetext}
The integral over \lz\ can be found analytically by means of an Euler 
substitution~\cite{gradstein}. The end result is
\begin{align}
	\label{eq:ifunctions}
	\mathfrak{I}_t (\en)= \frac{2}{\sqrt{\vqty{x-2}}} &
	\left[ \theta \pqty{x-x_{\text{mb}}^+}
	\pqty{\theta(x-2) \arctan \frac{1}{\sqrt{\kappa^+}} +
	\theta(2-x) \arctanh \frac{1}{\sqrt{-\kappa^+}}}\right. \notag\\
	+&\qquad \left.
	\theta \pqty{x-x_{\text{mb}}^-} \arctan \frac{1}{\sqrt{\kappa^- -1}}
	\right], \notag \\
	\mathfrak{I}_\phi(\en) = \frac{1}{\sqrt{\vqty{x-2}}} 
	& \left[ \theta \pqty{x-x_{\text{mb}}^+}
	\left(\theta(x-2) \Bqty{\pqty{\lplus+\lminus} 
	\arctan \frac{1}{\sqrt{\kappa^+}} + \sqrt{\pqty{\lplus-\lcritp}
	\pqty{\lcritp-\lminus}}} \right. \right. \notag \\ 
	&\qquad+
	\left.\theta(2-x)\Bqty{ \pqty{\lplus+\lminus} 
	\arctanh \frac{1}{\sqrt{-\kappa^+}} - \sqrt{\pqty{\lcritp-\lplus}
	\pqty{\lcritp-\lminus}}} \right) \notag\\
	&\qquad+\left.
	\theta \pqty{x-x_{\text{mb}}^-} 
	\Bqty{\pqty{\lplus+\lminus} \arctan \frac{1}{\sqrt{\kappa^- -1}} 
	-\sqrt{\pqty{\lplus-\lcritm}
	\pqty{\lcritm-\lminus}} }
	\right],
\end{align}
where we have defined
\begin{equation}
	\label{eq:kappa}
	\kappa^+ \equiv \frac{\lcritp - \lminus}{\lplus - \lcritp}
	\qquad
	\kappa^- \equiv \frac{\lplus - \lminus}{\lcritm-\lminus}.
\end{equation}
\end{widetext}

Since we are working with a geometry that has only two spatial dimensions,
our final result is a surface mass density. Working instead with the full
Kerr geometry and the restriction $\flz = \fplane 
\delta\pqty{u_\theta}$ on the distribution function, results in a 
volume mass density, which is what we used in \sect{constf}.
The only difference between the results of this appendix and those in
\sect{constf} is a factor of $2/r$ between the currents in \eq{current2d} and 
those in \eq{current_equator}. When calculating quantities integrated over
space, such as masses or annihilation fluxes, the difference is only a 
factor of $2$, since the three-dimensional Jacobian has an extra 
factor of $r$ compared to the planar geometry.

\bibliography{bhreferences}

\end{document}